\documentclass[prl,twocolumn,aps,showpacs]{revtex4}
\usepackage{graphicx}
\graphicspath{{pict/}{}}
\usepackage{color}

\newcommand{\be}{\begin{equation}}
\newcommand{\ee}{\end{equation}}

\newcounter{Fig}

\begin{document}

\title{Analytical theory for dark soliton interaction in nonlocal nonlinear materials with arbitrary degree of nonlocality}

\author{Qian Kong$^{1,2}$, Q. Wang$^{2}$, O. Bang$^3$, W. Krolikowski$^1$}

\address{$^1$Laser Physics Center, Research School of Physics and Engineering,
Australian National University, Canberra ACT 0200, Australia\\
$^2$Department of Physics, Shanghai University, Shanghai 200444, China\\
$^3$DTU Fotonik, Department of Photonics Engineering, Technical University of Denmark, 2800 Kgs. Lyngby, Denmark}




\begin{abstract}
We investigate theoretically the interaction of dark solitons in
materials with a spatially nonlocal nonlinearity. In particular we
do this analytically and for arbitrary degree of nonlocality. We
employ the variational technique to show that nonlocality induces
an attractive force in the otherwise repulsive soliton
interaction.
\end{abstract}

\pacs{42.65.Tg, 42.65.Sf, 42.70.Df, 03.75.Lm}
\maketitle

\section{Introduction}

Spatial optical solitons represent beams, which propagate in
nonlinear media without changing their profile. Their existence is
a result of an interplay  between size-determined diffraction and
nonlinearity-induced phase modulation, which in most cases is
produced by the refractive index modification of the material.
Depending on the type of nonlinearity, nonlinear media may support
either bright or dark solitons \cite{KivsharAgrawal}. While bright
solitons are just finite size beams formed in media with
self-focusing nonlinearity, dark solitons are more complex
objects, as they represent an intensity dip in an otherwise
constant background with nontrivial phase
profile~\cite{Kivshar:pr:98}. Spatial dark solitons have been
observed and studied in media with a negative or self-defocusing
nonlinearity \cite{Skinner:jqe:91,Swartzlander:prl:91}. Their
temporal counterparts, which have the form of "dark pulses", i.e.,
temporal intensity dips on a cw background, can exist in optical
fibers in the normal dispersion regime
\cite{Hasegawa:apl:73,Tomlinson:josab:89}. In recent years the
renewed interest in the properties of dark solitons stem from
experimental advances in the physics of matter waves. In
particular, the formation of dark matter wave solitons have been
observed in Bose Einstein condensates with repulsive
inter-particle interaction \cite{Burger:prl:99, Nath08:prl,
Stellmer:prl:08,Weller:prl:08}. There has also been report on the
possibility of dark soliton formation in nonlinear metamaterials
\cite{Weller:prl:08}. Interestingly, temporal dark solitons were
also shown to be able to induce supercontinuum generation in
photonic crystal fibers \cite{SCG_dark}.

The unique property of optical solitons, either bright or dark, is
their particle-like behavior in interaction \cite{KivsharAgrawal}.
However, it is also well known that there is a fundamental
difference in the interaction of bright and dark solitons. While
bright soltions may attract, repel, or even form bound states,
depending on their relative phase \cite{Gordon, Shalaby:ol:91,
Aitchison:ol:91}, dark solitons always repel. This has been
confirmed in numerous theoretical and experimental
works~\cite{Blow:pla:85, Zhao:ol:89, Forsua:prl:96}. We have shown
recently that the nature of dark soliton interaction can be
drastically altered by the spatially nonlocal character of
nonlinearity \cite{Nikolov:ol:04, Dreischuh:prl:06}. In nonlocal
media the nonlinear response of the medium in a particular spatial
location is determined not only by the wave (light) intensity in
that position, as in the local media, but also by the intensity in
a certain neighborhood around the point. As a result spatial
nonlocality provides stabilization of bright solitons
\cite{SnyderMitchell:science:97, Bang02, Skupin:pre:06}, and
induces their attraction, even if they are out-of-phase
\cite{Peccianti:ol:02, Rasmussen05}. Nonlocality has a similar
effect on dark solitons. In particular, it has been shown both
numerically  \cite{Nikolov:ol:04} and experimentally
\cite{Dreischuh:prl:06} that nonlocality induces attraction of
otherwise repelling dark solitons, leading to the formation of
their bound states. The physics of soliton attraction in nonlocal
nonlinear media can most easily be understood in the (linear)
regime of strong nonlocality \cite{SnyderMitchell:science:97}. In
the context of nonlinear optics a strongly nonlocal response of
the medium leads to the formation a broad (linear) index
waveguide, which can trap two or more solitons and enable the
formation of bound states. In the context of matter waves such
nonlocal (dipolar) interaction leads to the formation of a
potential well, which again induces attraction between solitons.
While dark soliton attraction has already been observed
experimentally \cite{Dreischuh:prl:06} the theoretical description
of this phenomenon in the regime of arbitrary degree of
nonlocality has been analyzed only numerically
\cite{Nikolov:ol:04} or in the special linear regime of strong
nonlocality \cite{Hu:apl:06}.

In this work we will investigate \emph{analytically} the
interaction of dark solitons in nonlocal media with an
\emph{arbitrary} degree of nonlocality. We will consider a
suitable nonlocal response function and use the variational
approach to derive analytical formulas for the forces acting
between two dark solitons. Our results clearly show how
nonlocality induces an attractive force, which depends on the
degree of nonlocality and counteracts the otherwise inherent
repulsive nature of dark soliton interaction.

\section{The nonlocal model and the response function}

In what follows we will be interested in the evolution of 1+1
dimensional optical beams with a scalar amplitude $E(x,z)$ and intensity $I(x,z)=|E(x,z)|^2$, that depends on the transverse $x$-coordinate and the propagation coordinate $z$. Propagation of such beams in materials with a
nonlocal defocusing nonlinearity can be modeled by the following
generic nonlocal nonlinear Schr\"{o}dinger (NLS) equation
\begin{equation}\label{NNLS}
i\frac{\partial E}{\partial z} + \frac{1}{2}\frac{\partial^{2}
E}{\partial x^{2}} - E\int^{+\infty}_{-\infty}
R(x-\xi)I(\xi,z)d\xi = 0,
\end{equation}
with the nonlocal response in the form of a convolution, where $R(x)$ is the nonlocal response function. In what follows we will use the normalization $\int_{-\infty}^{\infty}R(x)dx=1$. Obviously $R(x)=\delta(x)$ in a local Kerr medium.
The actual form of the nonlocal response is determined by the
details of the physical process responsible for the nonlocality. For
all diffusion-type nonlinearities \cite{Ghofraniha:prl:07}, orientational-type nonlinearities (like nematic liquid crystal) \cite{Peccianti:ol:02}, and for the general quadratic nonlinearity describing parametric interaction \cite{Nikolov03, Larsen06, Bache07, Bache08}, the response function is an exponential $R(x)=(2\sigma)^{-1}\exp(-|x|/\sigma)$ originating from a Lorentzian in the Fourier domain, with $\sigma >0$ defining the degree of nonlocality. Interestingly, for parametric interaction, the response function can also be periodic, $R(x)\propto\sin(|x|/\sigma)$ in certain regimes of the parameter space.

To obtain analytically tractable results the strongly nonlocal limit of $\sigma\rightarrow\infty$ is often used, in which the equation becomes linear \cite{Assanto:prl:04, Nikolov03, chinese_h_nonlocal, ShadrivovZharov} and the solitons are known as accessible solitons \cite{SnyderMitchell:science:97}. The so-called weakly nonlocal limit ($\sigma\ll1$) also presents a simpler model, which can be solved exactly for both dark and bright solitons \cite{wkob:pre:01}.

Other types of localized response functions have been used to obtain qualitative analytical results that captures the physics of the effect of nonlocality, such as a Gaussian in connection variational calculations \cite{who,Briedis:05}. The generic properties of the different types of response functions have been studied by Wyller et al. in terms of modulational instability and it was shown that in general all types of localized response functions have the same generic properties, provided their Fourier transform is positive definite \cite{Wyller02}.

Here we combine two approaches. First we use the weakly nonlocal model because it allows to study any localized response function by a single parameter. This allows us to derive the weakly nonlocal form of the interaction potential for any localized response function using the variational approach.
Then we introduce an arbitrary degree of nonlocality. We do this by assuming a box-type localized response function, because this allows us to calculate the integrals that appear in the variational approach. By comparing the results for arbitrary degree of nonlocality and the box-type response to the generic results obtained in the weakly nonlocal limit, we prove that the results are indeed generic.

\section{Interaction between dark solitons in weakly nonlocal medium}

We begin our analysis  by considering first the specific weakly
nonlocal limit of Eq.(\ref{NNLS}), in which the width of the response
function is much smaller than the spatial scale of the solitons.
Then the intensity of the beam $I(\xi,z)$ can be expanded in a
Taylor series with respect to $\xi$ around $\xi=x$, and Eq.(1)
turns into
\begin{equation} \label{weakNLS}
i\frac{\partial E}{\partial z} + \frac{1}{2}\frac{\partial^{2}
E}{\partial x^{2}} - E\left(I+\gamma\frac{\partial^{2}I}{\partial
x^{2}}\right) = 0,
\end{equation}
where $\gamma=\frac{1}{2}\int^{+\infty}_{-\infty}R(x)x^{2}dx$ clearly shows how the response function needs to be localized. It is important to note that we have here assumed a symmetric response function, which is why it is the second derivative that appears as the perturbation term proportional to $\gamma$. Asymmetric response functions, such as the Raman response in optical fibers, could of course easily be used too. However, asymmetric response functions do not allow for defining a Lagrangian and thus to use the variational approach. Thus we consider here only symmetric response functions.

We will investigate the dark solitons using the variational (or Lagrangian)
approach~\cite{Anderson:pra:83}. It can be shown that the
Lagrangian density corresponding to Eq.(\ref{weakNLS}) is of the following form
\begin{eqnarray}
\mathcal{L}&=&\frac{i}{2}\left(u^{*}\frac{\partial u}{\partial
z}-u\frac{\partial u^{*}}{\partial
z}\right)\left(1-\frac{1}{|u|^{2}}\right)-\frac{1}{2}\left|\frac{\partial
u}{\partial
x}\right|^{2}\nonumber \\
&-&\frac{1}{2}(|u|^{2}-1)^{2}+\frac{1}{2}\gamma\left(\frac{\partial
|u|^{2}}{\partial x}\right)^{2},
\end{eqnarray}
where we normalized the background intensity of the solitons to unity and
used the following transformation for the amplitude of the field $E(x,z)=u(x,z)\exp(iz)$.
To proceed further we must postulate the form of the function $u(x)$.
It was already shown earlier in studies of local dark solitons that
the proper ansatz  is of the form
\begin{equation}
u = (B \tanh z_{+}-i A)(B \tanh z_{-}+i A),
\end{equation}
where $z_{\pm} = D(x\pm x_{0})$ and $2x_0$ denotes the separation between  solitons and $A,B$ satisfy the normalization condition $A^2+B^2=1$.
The choice of this particular ansatz is dictated by the fact that it represents {\em exact}
dark soliton solutions of noninteracting local dark solitons.
Substituting Eq.(4) into Eq.(3) and considering the case of weakly
overlapping dark solitons, we obtain the averaged Lagrangian
$L=\int_{-\infty}^{\infty}\mathcal{L}dx$  in the following form

\begin{eqnarray}
&L &=
2L_{0}+\frac{dB}{dz}\frac{4B^{2}}{AD\tanh(2x_{0}D)}+16B^{2}\mbox{e}^{-4x_{0}D}\nonumber \\
&\times &\left[2x_{0}(D^{2}-B^{2}-4\gamma
B^{2}D^{2})+\frac{B^{2}}{3D}(4B^{2}-D^{2})\right].
\end{eqnarray}
where
\begin{eqnarray}\label{lagrangian}
L_{0}&=&2\frac{dx_{0}}{dz}\left[-AB+\tan^{-1}\left(\frac{B}{A}\right)\right]\nonumber \\
&-&\frac{2}{3}\left[B^{2}D+\frac{B^{4}}{D}\right]+\frac{8}{15}\gamma
B^{4}D.
\end{eqnarray}
 is the Lagrangian for the noninteracting weakly nonlocal dark solitons~\cite{Wang:jopt:09}.

From the corresponding Euler-Lagrangian equations one finds the
following relations for soliton parameters
\begin{eqnarray}
\frac{dA}{dz}&=&8BD\mbox{e}^{-4x_{0}D}\left[\frac{}{}2x_{0}(D^{2}-B^{2}
- 4\gamma B^{2}D^{2})\right. \nonumber \\
&+&\left.\frac{B^{2}}{3D}(4B^{2}-D^{2})\right].
\end{eqnarray}

\begin{eqnarray}
& &\frac{1}{3}\left[1-\frac{B^{2}}{D^{2}}\right]
+16x_{0}\mbox{e}^{-4x_{0}D}\left[\frac{}{} 2x_{0}(D^2-B^{2}-4\gamma
B^{2})\right.\nonumber \\
&+&\left.\frac{B^{2}}{3D}(4B^{2}-D^2)\right] -\frac{4}{15}\gamma
B^{2}=0.
\end{eqnarray}

\begin{eqnarray}
\hspace{-10mm}&&\frac{dx_{0}}{dz}-\frac{A}{3}\left[\frac{D}{B}+\frac{2B}{D}-\frac{8}{5}\gamma
BD\right]+\left[\frac{}{}x_{0}(D^{2}-2B^{2}\right.\nonumber\\
&-&\left.8\gamma
B^{2}D^{2})+\frac{B^{2}}{3D}(6B^{2}-D^{2})\right]\frac{8A}{B}\mbox{e}^{-4x_{0}D} \nonumber \\
&+&\frac{dD}{dz}[2D^{2}\tanh(2x_{0}D)]^{-1}=0.
\end{eqnarray}

Assuming well separated and weakly interacting ($x_0D\gg1$) almost "black" solitons $(A^{2}\approx 0)$
we can obtain from Eqs.~(7-9) the following equation for the soliton coordinate $x_0$
\begin{eqnarray}
\frac{d^{2}x_{0}}{dz^{2}}= -\frac{dV(x_{0})}{dx_{0}},
\end{eqnarray}
\noindent
where we have introduced the "potential" $V(x_0)$ as
\begin{equation}
V(x_{0})=V_1(x_0)+V_2(x_0)
\end{equation}
with
\begin{eqnarray}
&V_1(x_{0})&=\frac{2\left(1-\frac{16}{15}\gamma
B^{2}\right)}{1-\frac{4}{5}\gamma
B^{2}}\exp\left (-\frac{4x_{0}B}{\sqrt{1-\frac{4}{5}\gamma  B^{2}}}\right )B^4\nonumber \\
&V_2(x_{0})&=-\frac{2\left(1-\frac{16}{15}\gamma
B^{2}\right)}{1-\frac{4}{5}\gamma
B^{2}}\exp\left (-\frac{4x_{0}B}{\sqrt{1-\frac{4}{5}\gamma  B^{2}}}\right )\nonumber \\
&\times & \hspace{-5mm}B^{4}\left[\frac{8}{5}\gamma
\left(\frac{4x_{0}B}{\sqrt{1-\frac{4}{5}\gamma
B^{2}}}+1+\frac{2}{3}B^2\right)\right].
\end{eqnarray}
Therefore, for set values of the soliton parameter $B$ and nonlocality
 $\gamma$, the dynamics of soliton interaction is represented as a
mechanical analogy describing the motion of a particle in an external
potential. The potential consists of two contributions. The first one , $V_1(x_0)$, which exists even for local nonlinearity, is positive and hence is responsible for the naturally occurring dark soliton repulsion \cite{Zhao:ol:89,Kivshar:oc:95,Theocharis:pra:05}. The second contribution, $V_2(x_0)$, provides a nonlocality-mediated attractive force, which disappears for $\gamma=0$.

The simultaneous presence of competing repulsive and nonlocality-induced attractive forces introduces a local well in the soliton interaction potential $V(x_0)$, as clearly demonstrated in Fig.~1 for $\gamma$=0.05, which enables the formation of soliton bound states otherwise not possible in the local NLS equation. This result is obtained on the specific weakly nonlocal limit, which has the nice advantage of being generic, in the sense that it is \emph{valid for any localized and symmetric response function}. In the following section we will extend the results to the full regime of an arbitrary degree of nonlocality by considering a specific response function. However, we connect the general results to the generic result of the weakly nonlocal limit to demonstrate the generic nature of also the general result.

\begin{figure}
\includegraphics[width=1.0\columnwidth]{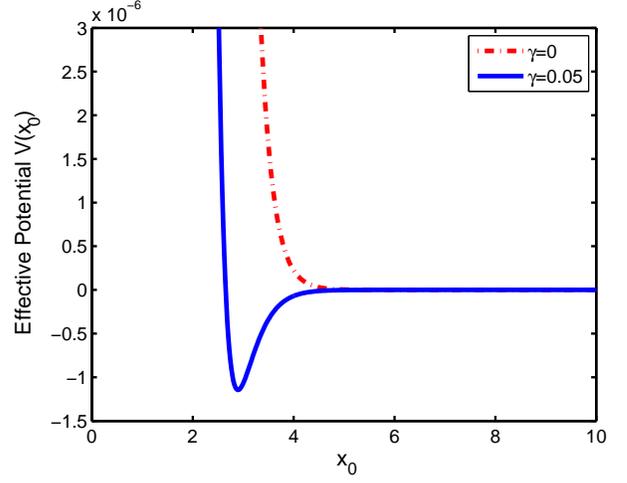}
\caption{\label{fig1}Dark soliton weakly nonlocal interaction
potential $V(x_{0})$, given by Eq.(11). Red dashed line - local
case ($\gamma=0$); blue solid line - weakly nonlocal regime
($\gamma=0.05$).}
\end{figure}

\section{General nonlocal case}
Here we consider the interaction between the dark solitons in
nonlocal media, in which the nonlocal response has an arbitrary
degree of nonlocality. Then the Lagrangian density corresponding
to Eq.(1) is

\begin{eqnarray}
&\mathcal{L}&=\frac{i}{2}\left(u^{*}\frac{\partial u}{\partial
z}-u\frac{\partial u^{*}}{\partial
z}\right)\left(1-\frac{1}{|u|^{2}}\right)
-\frac{1}{2}\left|\frac{\partial
u}{\partial x}\right|^{2}\nonumber \\
\hspace{-2mm}&-&\frac{1}{2}\left(|u|^{2}-1\right)\int^{+\infty}_{-\infty}
\hspace{-2mm}R(x-\xi)\left(|u(\xi,z)|^{2}-1\right)d\xi.
\end{eqnarray}

In order to make the problem analytically tractable we will
consider here the particular model of nonlocality described by the
rectangular nonlocal response function,

\begin{equation}
R(x)=\left\{\begin{array}{ll} \frac{1}{2\sigma} & -\sigma \leq x \leq \sigma,\\
&\\
0 & \mbox{otherwise.}\end{array}\right.
\end{equation}
Physically, this type of nonlocal response means that the nonlinear response of the medium
in a particular spatial location is determined by the equal contributions from the light intensity
in the neighborhood of this location defined by parameter $\sigma$. This is obviously a simplification, but
as we will see later, it leads to a physically correct description of the soliton interaction.

Substituting Eq.(4) into Eq.(12) and integrating over transverse
coordinate  $x$, we obtain the averaged Lagrangian in the form

\begin{eqnarray}
L &=& \frac{dB}{dz}\frac{4B^{2}}{AD\tanh(2x_{0}D)}
+4\frac{dx_{0}}{dz}\left[-AB+\tan^{-1}\left(\frac{B}{A}\right)\right]\nonumber
\\
&-&\frac{4}{3}B^{2}D\left[4B^{2}+12-24Dx_{0}\right]\mbox{e}^{-4x_{0}D}-\frac{4}{3}B^{2}D \nonumber \\
&+&\frac{2B^{4}}{D}\left[\mbox{csch}^{2}(D\sigma)-\frac{\coth(D\sigma)}{D\sigma}\right]
+\frac{4B^{4}}{D}\left[2\cosh(2D\sigma)\right. \nonumber \\
&-&\left.(4Dx_{0}-1)\frac{\sinh(2D\sigma)}{D\sigma}+8B^{2}+4B^{2}\mbox{csch}^{2}(D\sigma)\right.\nonumber \\
&-&\left.\frac{4B^{2}\coth(D\sigma)}{D\sigma}\right]\mbox{e}^{-4x_{0}D},
\end{eqnarray}

From the corresponding Euler-Lagrangian equations one can derive
 the evolution equation for the soliton coordinate,
 which in the limit of weakly interacting (i.e. well separated), almost black solitons $(A^{2}\ll 1)$
takes the following form

\begin{eqnarray}
& &\frac{d^{2}x_{0}}{dz^{2}}=\left[\frac{D}{3B}-\frac{B}{D}\left(\mbox{csch}^{2}(D\sigma)
-\frac{\coth(D\sigma)}{D\sigma}\right)\right]\nonumber \\
&\times&\left \{\frac{2}{3}BD^{2}\left(24Dx_{0}-4B^{2}-12\right)\right.\nonumber
\\
&+&\left.2B^{3}\left[(1-4Dx_{0})\frac{\sinh(2D\sigma)}{D\sigma}+2\cosh(2D\sigma)\right.\right.\nonumber \\
&+&\left.\left.8B^{2}+4B^{2}\mbox{csch}^{2}(D\sigma)-\frac{4B^{2}\coth(D\sigma)}{D\sigma}\right]\right\}
\mbox{e}^{-4x_{0}D} \nonumber \\
 &=&-\frac{dV(x_{0})}{dx_{0}},
\end{eqnarray}
where the effective potential function $V(x_{0})$ is defined as

\begin{eqnarray}\label{potential_general}
&&V(x_{0})=\left[\frac{D^2}{3}-B^2\left(\mbox{csch}^{2}(D\sigma)-\frac{\coth(D\sigma)}{D\sigma}\right)\right]
\nonumber \\
& & \left\{(1+4Dx_{0})\left(1-\frac{B^{2}}{2D^2}\frac{\sinh(2D\sigma)}{D\sigma}\right)-\frac{2}{3}(B^{2}+3)\right.\nonumber
\\
&+&\left.\frac{B^{2}}{2D^2}\left[\frac{\sinh(2D\sigma)}{D\sigma}+2\cosh(2D\sigma)+8B^{2}\right.\right.\nonumber \\
&+&\left.\left.4B^{2}\mbox{csch}^{2}(D\sigma)-\frac{4B^{2}\coth(D\sigma)}{D\sigma}\right]\right\}\mbox{e}^{-4x_{0}D},
\end{eqnarray}
and parameters $B$, $D$ and $\sigma$ satisfy the following relation
\begin{equation}
 \frac{1}{3}-\frac{B^{2}}{D^{2}}\frac{\coth(D\sigma)}{D\sigma}\left[1-D^{2}\sigma^{2}\mbox{csch}^{2}(D\sigma)\right]=0.
\end{equation}

\begin{figure}
\includegraphics[width=1.0\columnwidth]{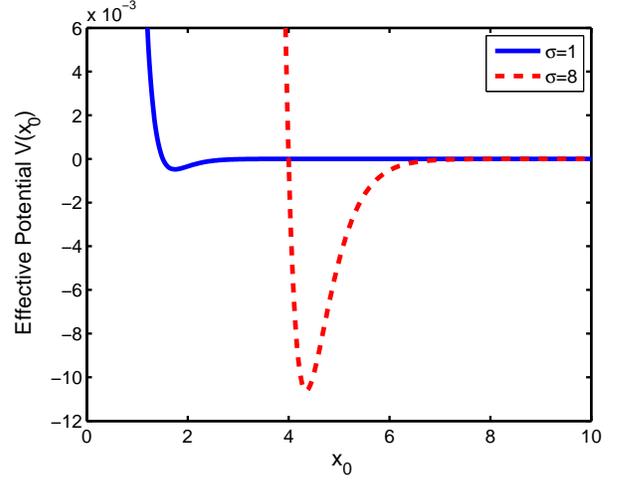}
\caption{\label{fig2}Dark soliton interaction potential
$V(x_{0})$, given by Eq.(16), for a rectangular nonlocal response
with an arbitrary degree of nonlocality $\sigma$. Red dashed line
- ($\sigma=1$); blue solid line - strongly nonlocal regime
($\sigma=8.0$).}
\end{figure}

One can show that in the weakly nonlocal limit, i.e., when
$\sigma\ll1$ the formula Eq.~(16) leads to the potential of the
form of Eq.~(11) with the nonlocality parameter $\gamma$ given by
$\gamma=\frac{1}{2}\int^{+\infty}_{-\infty}R(x)x^{2}dx=\sigma^{2}/6$.
In Fig.~2 we show the potential $V(x_0)$ for different values
of the nonlocality $\sigma$.  It is evident that the generic results of the weakly
nonlocal model remain valid also for an arbitrary degree of nonlocality, i.e., nonlocality
provides an attractive contribution to the potential, which
counteracts the natural repulsion of dark solitons thus enabling
the formation of their bound states.
This fact provides evidence that our general results for the specific rectangular response
 function are, in fact,  generic also for any symmetric and localized response function.

We now confirm our theory by direct numerical simulations of the nonlocal NLS Eq.(1) with rectangular nonlocal response function.  As initial conditions we used Kerr soliton profiles (see Eq.(4)) with $A=0$, $B=D=1$.
The representative results are depicted in Fig.3. These
contour plots show the dynamics of initially well separated
solitons. The separation is chosen in such a way that both solitons clearly repel
when the nonlinearity is local (Fig.3(a)). It is clear that as the extent of nonlocal
response increases both solitons  start experiencing the  attractive force.
 In fact, in  case depicted in Fig.3(b) ($\sigma=1$)
  the natural repulsion of solitons is almost completely compensated for by the nonlocality-mediated attraction
  leading to the formation of the bound state of dark solitons. Interestingly, in this case both solitons are separated by the distance of 2$x_0$=3.8 which corresponds to the location of the minimum of the effective potential  from Fig.2 for $\sigma$=1.
  For even stronger nonlocality the attractive force causes  mutual oscillations of  solitons trajectories.  The radiation visible in Fig.3(b-d) is a result of the fact that the initial wave profiles are not exact dark solitons in the nonlocal regime. Hence, the solitons  evolve and transform as they propagate shading away radiation.
\begin{figure}\label{fig3}
\includegraphics[width=1.0\columnwidth]{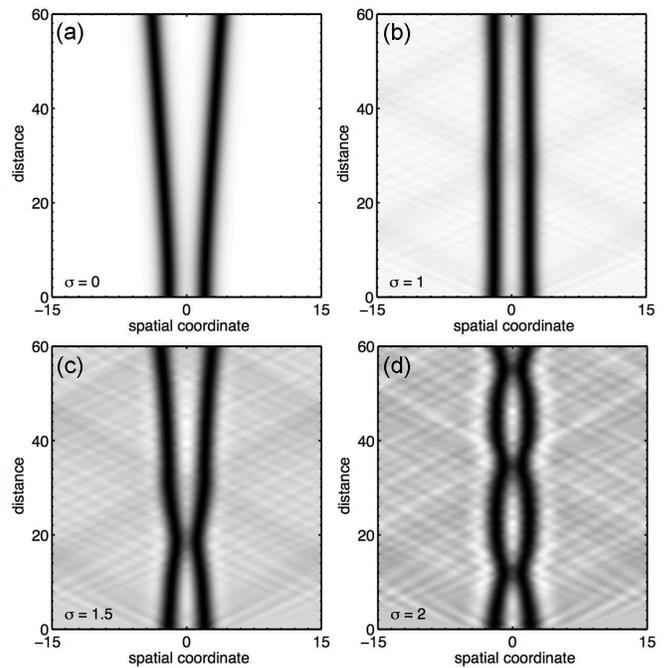}
\caption{Numerically simulated interaction of initially parallel
dark solitons in nonlocal medium with rectangular nonlocal
response function, for different degree of nonlocality  (a)
$\sigma=0$; (b) $\sigma=1$, (c) $\sigma=1.5$, (d) $\sigma=2.0$.
Notice the almost parallel propagation of solitons for $\sigma=1$
indicating the balance between repulsive and attractive forces
and, consequently, formation of the soliton bound state.}
\end{figure}

In Fig.4 we plot with the solid line the  separation between solitons corresponding to their bound state as calculated from the minimum of the effective potential Eq.(\ref{potential_general}). It is evident the separation is nonmonotonic function of the degree of nonlocality. This can be explained as follows. For small $\sigma$ the nonlocality-mediated attractive forces are very weak. Therefore the only way to  compensate the natural repulsion of the solitons is to increase their separation until the latter sufficiently decreases. On the other hand, for large $\sigma$ the nonlocal nonlinear potential becomes very broad resulting again in an increased  separation of solitons. This behavior has been confirmed in numerical simulations. To this end, for given degree of nonlocality  we varied the initial distance between the solitons  and numerically propagated them over distance long enough to establish the formation of their bound state. The resulting separation is depicted in Fig.4 by filled squares. Clearly, it follows the trend found from variational analysis. On the other hand, the numerical data is limited to relatively low degree of nonlocality because the strong  radiation for larger  $\sigma$ prevents the accurate determination of the bound states.

\begin{figure}\label{fig4}
\includegraphics[width=1.0\columnwidth]{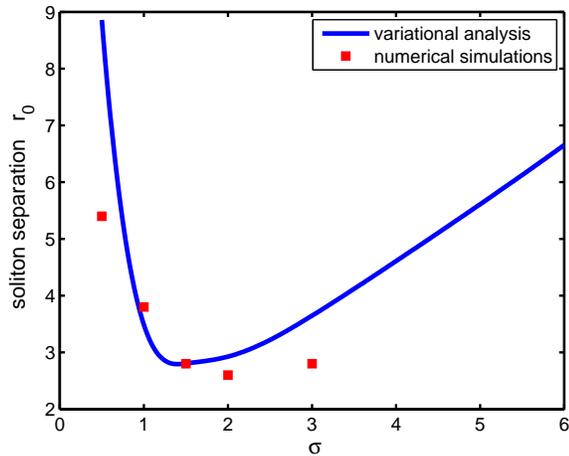}
\caption{Separation between solitons in a bound state ($r_0$) as a
function of the degree of nonlocality $\sigma$. Solid line -
variational calculations; squares - numerical simulations of
soliton propagation.}
\end{figure}
\section{Conclusion}
We studied analytically the interaction of dark spatial solitons in nonlocal medium.
 We used  variational technique to derive the evolution equations for
 the separation between both solitons.
We showed that nonlocality provides an attractive force between otherwise
repelling solitons. For high degree of  nonlocality the attractive force is strong enough to enable
 formation of bound states of dark solitons.

\section{Acknowledgement}

This work was supported by the National Natural Science Foundation
of China (Grant No. 60808002), the Shanghai Leading Academic
Discipline Program (Grant No. S30105), the China Scholarship
Council and the Australian Research Council.


\begin{thebibliography}{99}

\bibitem{KivsharAgrawal} Yu.~S. Kivshar, G. Agrawal,
{\em Optical Solitons: From Fibers to Photonic Crystals} (Academic
Press, San Diego, 2003).

\bibitem{Kivshar:pr:98} Y.S.~Kivshar  and  B.~Luther-Davies, Phys. Rep. {\bf 298}, 81 (1998)
   and references therein.

\bibitem{Skinner:jqe:91} S.~R. Skinner, G.~R. Allan, D.~R. Andersen, and A.~L.
Smirl, IEEE J. Quantum Electron. {\bf 27,} 2211--2219 (1991).

\bibitem{Swartzlander:prl:91}G. Swartzlander, D. R. Andersen, J. J. Regan, H. Yin,
and A. E. Kaplan, Phys. Rev. Lett. {\bf 66}, 1583--1586 (1991).


\bibitem{Hasegawa:apl:73}A. Hasegawa and F. Tappert, Appl. Phys. Lett. {\bf 23}, 171-172 (1973).

\bibitem{Tomlinson:josab:89}W.~J. Tomlinson, R.~J. Hawkins, A.~M. Weiner, J.~P. Heritage, R.~N. Thurston, J. Opt. Soc. Am. {\bf 6}, 329--334  (1989).

\bibitem{Burger:prl:99}S. Burger, K. Bongs, S. Dettmer, W. Ertmer, and K. Sengstock, A. Sanpera1,
G. V. Shlyapnikov, and M. Lewenstein, Phys. Rev. Lett. {\bf 83},
5198–-5201 (1999).

\bibitem{Nath08:prl}R.~Nath, P.~Pedri, and L.~Santos, Phys. Rev. Lett. {\bf 101}, 210402 (2008).

\bibitem{Stellmer:prl:08}S. Stellmer, C. Becker, P. Soltan-Panahi,
 E.-M. Richter, S. Dorscher, M. Baumert, J. Kronjager,
 K. Bongs, K. Sengstock, Phys. Rev. Lett. {\bf 101}, 120406--120409 (2008).

\bibitem{Weller:prl:08}A. Weller, J.P. Ronzheimer, C. Gross, J. Esteve, M.~K. Oberthaler,
D.~J. Frantzeskakis, G.  Theocharis, P.~G. Kevrekidis, Phys. Rev.
Lett. {\bf 101},  130401--130404   (2008).

\bibitem{SCG_dark}C. Mili$\acute{a}$n, D.~V. Skryabin, and A. Ferrando, Opt. Lett. {\bf 34}, 2096--2098 (2009).


\bibitem{Gordon}J.~P. Gordon, Opt. Lett. {\bf 8}, 596--598  (1983).

\bibitem{Shalaby:ol:91}M. Shalaby, A. Barthelemy, Opt. Lett. {\bf 16}, 1472--1474 (1991).

\bibitem{Aitchison:ol:91}J.~S. Aitchison, A.~M. Weiner, Y. Silberberg, D.~E. Leaird, M.~K. Oliver, J.~L. Jackel, P.~W.~.E. Smith, Opt. Lett. {\bf 16}, 15--17 (1991).

\bibitem{Blow:pla:85}K.~J. Blow and N.~J. Doran, Phys. Lett. A {\bf 107}, 55--58 (1985).

\bibitem{Zhao:ol:89} W. Zhao and E. Bourkoff, Opt. Lett. {\bf 14}, 1371--1373 (1989).

\bibitem{Forsua:prl:96}D. Foursa and P. Emplit, Phys. Rev. Lett. {\bf 77}, 4011--4014 (1996).

\bibitem{Nikolov:ol:04}N. Nikolov, W. Krolikowski, O. Bang, J.~J. Rasmussen, P.~L. Christiansen, M.~K. Oberthaler, Opt. Lett. {\bf 29}, 286--288 (2004).

\bibitem{Dreischuh:prl:06}A. Dreischuh, D.~N. Neshev, D.~E. Petersen, O. Bang, and W. Krolikowski, Phys. Rev.
Lett. {\bf 96}, 043901--043904 (2006).



\bibitem{SnyderMitchell:science:97}A. Snyder and J. Mitchell, Science {\bf 276},  1538 (1997).

\bibitem{Skupin:pre:06}S. Skupin, O. Bang, D. Edmundson, and W. Krolikowski, Phys. Rev. E {\bf 73}, 066603 (2006).


\bibitem{Bang02}
O. Bang, W. Krolikowski, J. Wyller, and J.J. Rasmussen, Phys. Rev.
E {\bf 66}, 046619 (2002).


\bibitem{Peccianti:ol:02}M. Peccianti, K.~A. Brzdakiewicz, and G. Assanto, Opt. Lett. {\bf 27}, 1460-1462 (2002).

\bibitem{Rasmussen05}
P.D. Rasmussen, O. Bang, and W. Krolikowski, Phys. Rev. E {\bf
72}, 066611 (2005).



\bibitem{Hu:apl:06}W. Hu, T. Zhang, Q. Guo, L. Xuan, and S. Lan, Appl. Phys. Lett. {\bf 89}, 071111 (2006).

\bibitem{Ghofraniha:prl:07}N. Ghofraniha, C. Conti, G. Ruocco, and S. Trillo, Phys. Rev. Lett. {\bf 99}, 043903 (2007).

\bibitem{Nikolov03}
N.I. Nikolov, D. Neshev, O. Bang, and W. Krolikowski, Phys. Rev.
E {\bf 68}, 036614 (2003).

\bibitem{Larsen06}
P.V. Larsen, M.P. Sørensen, O. Bang, W.Z. Krolikowski, S. Trillo,
Phys. Rev. E {\bf 73}, 036614 (2006).

\bibitem{Bache07}
M. Bache, O. Bang, J. Moses, F.W. Wise, Opt. Lett. {\bf 32}, 2490
(2007).


\bibitem{Bache08}
M. Bache, O. Bang, W. Krolikowski, J. Moses, F.W. Wise, Opt.
Express {\bf 16}, 3273-3287 (2008).


\bibitem{Assanto:prl:04}C. Conti, M. Peccianti, G. Assanto, Phys. Rev. Lett. {\bf 92}, 113902 (2004).

\bibitem{chinese_h_nonlocal}D. Deng, Q. Guo, and W. Hu, Phys. Rev. A {\bf 79}, 023803 (2009);
W.-P. Zhong and M. Belic, Phys. Rev. A {\bf 79}, 023804 (2009); D.
Deng, Q. Guo, Journal of Optics A Pure and Applied Optics {\bf
10}, 035101 (2008).

\bibitem{ShadrivovZharov}I.~V. Shadrivov and A. A. Zharov, J. Opt. Soc. Am. B {\bf 19}, 596--602 (2002).

\bibitem{wkob:pre:01}W. Krolikowski and O. Bang, Phys. Rev. E {\bf 63}, 016610--016615 (2001).

\bibitem{who} M. Shen, N. Xi, Q. Kong, L-J. Ge, J-L.  Shi and Q. Wang, Chinese Phys. B {\bf 18}, 2822 (2009).

\bibitem{Briedis:05}D. Briedis, D. Edmundson, O. Bang, and W. Krolikowski, Opt. Express {\bf 13}, 435--443 (2005);A. I. Yakimenko, V. M. Lashkin, and O. O. Prikhodko, Phys. Rev. E {\bf 73}, 066605 (2006); S. Skupin, M. Grech, and W.Krolikowski, Opt. Express. {\bf 16,} 9118--9131 (2008).

\bibitem{Wyller02}
J. Wyller, W. Krolikowski, J.J. Rasmussen, Phys. Rev. E {\bf 66},
066615 (2002).

\bibitem{Anderson:pra:83}D.Anderson, Phys. Rev. A {\bf 27}, 3135 (1983).








\bibitem{Wang:jopt:09}L.~J. Ge, Q. Wang, M. Shen, J. Shi, Q. Kong and P. Hou, J.
Opt. A {\bf 11}, 065207  (2009).

\bibitem{Kivshar:oc:95}Yu.~S. Kivshar, W. Krolikowski, Opt. Commun. {\bf 114,} 353--362 (1995).

\bibitem{Theocharis:pra:05}G. Theocharis, P. Schmelcher, M.~K. Oberthaler, P.~G. Kevrekidis, D.~J. Frantzeskakis,
Phys. Rev. A    {\bf 72}, 023609 (2005).

























\end{thebibliography}
\end{document}